# Different mechanics of snap-trapping in the two closely related carnivorous plants *Dionaea muscipula* and *Aldrovanda vesiculosa*

Simon Poppinga[1] and Marc Joyeux[2,#]

[1] *Plant Biomechanics Group Freiburg, Botanic Garden, Faculty of Biology, University of Freiburg, Schänzlestrasse 1, 79104 Freiburg im Breisgau, Germany*

[2] *Laboratoire Interdisciplinaire de Physique (CNRS UMR 5588), Université Joseph Fourier Grenoble 1, BP 87, 38402 St Martin d'Hères, France*

**Abstract**: The carnivorous aquatic Waterwheel Plant (*Aldrovanda vesiculosa* L.) and the closely related terrestrial Venus Flytrap (*Dionaea muscipula* SOL. EX J. ELLIS) both feature elaborate snap-traps, which shut after reception of an external mechanical stimulus by prey animals. Traditionally, *Aldrovanda* is considered as a miniature, aquatic *Dionaea*, an assumption which was already established by Charles Darwin. However, videos of snapping traps from both species suggest completely different closure mechanisms. Indeed, the well-described snapping mechanism in *Dionaea* comprises abrupt curvature inversion of the two trap lobes, while the closing movement in *Aldrovanda* involves deformation of the trap midrib but not of the lobes, which do not change curvature. In this paper, we present the first detailed mechanical models for these plants, which are based on the theory of thin solid membranes and explain this difference by showing that the fast snapping of *Aldrovanda* is due to kinematic amplification of the bending deformation of the midrib, while that of *Dionaea* unambiguously relies on the buckling instability that affects the two lobes.

## I - INTRODUCTION

Although having no muscles, some plants and fungi are able to perform very fast movements enabling them to achieve essential functions. The physical mechanisms involved in these movements differ and depend on the dimensions of the organism, the physical environment (e.g. density of the medium) and the characteristic movement speed that must be achieved [1]. The fastest motions are obtained via explosive dehiscence. For example, the sporangium discharge of the fungus *Pilobolus* lasts about 10 μs, while *Hura crepitans*, the Sandbox Tree, can fling seeds as far as 100 m away with a discharge time of about 100 μs. These mechanisms are exceedingly fast, but take place only once because plant tissues are torn during the process. On the other hand, fast and repetitive movements in plants are often due to rapid geometric changes of specific organs associated with the buckling of a thin membrane. For example, the millimeter-sized underwater suction traps of carnivorous *Utricularia* species (Bladderworts) catch their prey in a few ms, owing to the ability of the trap door to perform more than 100 rapid buckling/unbuckling cycles during the lifetime of the trap [2-4].

At last, the repetitive movements of larger plant organs, which are usually substantially less rapid, rely essentially on cell swelling/shrinking mechanisms and not on dehiscence or buckling. Many of them are turgor-dependent, which is the cell sap pressure acting against and deforming the cell walls. It is determined by vacuolar water content resulting from osmotic pressure, which may reach 1 MPa. Turgor pressure variations in so-called motor cells can actuate organ movement, as single motor cells may undergo a 25% volume change within 1 s and hence can effect cell stiffness in short time [5]. Groups of antagonistic motor cells in "hinges" (pulvini) lose turgor pressure upon stimulation (extensors) or become consequentially stretched (flexors), affecting the organ's bending stiffness and actuating the movement (e.g. folding). Based on this mechanism, the centimeter-sized leaves and leaflets of *Mimosa pudica,* for example, can fold in about 1 s after reception of an external stimulus.

However, it is often difficult to determine with certainty which mechanism originally initiates and causes a rapid nastic plant motion. For example, *Dionaea muscipula* is a small perennial herb from North America. Its leaves can clearly be divided into a lower part for photosynthesis (the petiole) and an upper part for prey capture (the leaf lamina). The upper part, here referred to as the trap, has a typical size in the range of 2-6 cm and consists of a pair of trapezoidal lobes held together along a midrib. These lobes snap in about 0.1-0.7 s when one of the three trigger hairs located at the center of each lobe is stimulated. Although this rapid closure has been known for more than a century (Darwin called *Dionaea* "one of the most wonderful plants in the world" [6]), there is still no general agreement concerning its mechanism. Proposed explanations are an irreversible acid-induced wall loosening [7] or a rapid loss of turgor pressure in motor cells [8], but it has been pointed out that none of these cellular mechanisms can account for the rapidity of the closure [9,10]. It was consequently suggested that elastic deformations and buckling may play an important role [9,11], but the need for buckling has been recently questioned [12].

On the other hand, *Aldrovanda vesiculosa* is a rootless, submerged aquatic herb with an almost worldwide distribution, which develops whorls of 7 or 8 leaves per node. Each leaf features a pair of oval lobes that are 4 to 7 mm long. The lobes shut in about 100 ms upon excitation of one of the 20 sensitive hairs located on the inner surface of each lobe. Because of the skills required to grow it, as well as the small size and speed of its traps, *Aldrovanda* has been much less studied than *Dionaea*. This is probably the reason why, following Darwin, it has repeatedly been described as "a miniature, aquatic *Dionaea*" [6]. Still, while phylogenetic studies confirm that *Dionaea* and *Aldrovanda* are sister species [13], previous classification of plant and fungal movements have classified *Dionaea* in the plant set of snap-buckling dominated movements but *Aldrovanda* in the set of swelling/shrinking dominated ones [1,14].

In this paper, we demonstrate that the establishment of detailed models based on the theory of thin solid membranes, in conjunction with high-speed camera recordings, may provide an unambiguous characterization of the physics involved in such movements. More precisely, we derive such models for the two closely related carnivorous plant species, the Venus Flytrap (*Dionaea muscipula*) and the Waterwheel Plant (*Aldrovanda vesiculosa*), and show that these sister species from the Sundew family (Droseraceae) actually use rather different mechanisms to achieve their startling capture speeds.

The the paper is organized as follows. The general features of the membrane model and the evolution equations are sketched in Sec. II. We then discuss in detail the snap-trapping mechanisms of *Aldrovanda* and *Dionaea* in Secs. III and IV, respectively. In these sections, special attention is paid to the description of how differences in turgor pressures are modeled, which is one of the central and most original points of this work. We finally conclude in Sec. V.

## II - THE MEMBRANE MODEL

We developed detailed mechanical models, which enable to solve most of the questions raised in the Introduction. These models rely on the fact that the thickness $h$ of the leaves is small compared to their width and length. $h$ is indeed of the order of 400 μm for *Dionaea* and in the range 40-70 μm for *Aldrovanda*. Leaves can therefore be described as thin solid membranes with total elastic energy $E_{pot} = E_{strain} + E_{curv}$, where

$$E_{\text{strain}} = \frac{E\,h}{2(1-\nu^2)} \int_S [(1-\nu)\text{Tr}(\boldsymbol{\varepsilon}^2) + \nu(\text{Tr}(\boldsymbol{\varepsilon}))^2]\,dS$$

$$E_{\text{curv}} = \frac{E\,h^3}{24(1-\nu^2)} \int_S [(\text{Tr}(\mathbf{b}))^2 - 2(1-\nu)\text{Det}(\mathbf{b})]\,dS\;.$$

(2.1)

$E_{\text{strain}}$ and $E_{\text{curv}}$ are the strain and curvature contributions arising from in-plane and out-of-plane deformations, respectively [15,16]. In this equation, $E$ stands for the Young's modulus of elasticity of the leaf, $\nu$ for its Poisson's ratio, $\mathbf{b}$ for the difference between the strained and unstrained local curvature tensors, $\boldsymbol{\varepsilon}$ for the two-dimensional Cauchy-Green local strain tensor, and integration is performed over the surface $S$ of the leaf. For a given leaf geometry, this model essentially contains no adjustable parameter. Indeed, the leaf tissue is almost incompressible, which implies that $\nu = 1/2$. Moreover, $E$ contributes only as a multiplicative term to the total elastic energy of the leaf. It therefore plays no role in the static mechanics of the plant organ. One can consequently learn a lot by building a surface with lifelike geometry, applying appropriate constraints to this surface and computing the response of the leaf, either by minimizing $E_{\text{pot}}$ or by integrating Langevin equations of the motion. This proved to be an efficient strategy to unravel the buckling/unbuckling mechanism at the origin of the ultra-fast and repetitive opening of *Utricularia* trapdoors [2,3]. However, there exists one fundamental difference, in the sense that constraints in *Utricularia* arise from the pressure difference between the liquid inside and outside of the trap. In contrast, for both *Dionaea* and *Aldrovanda*, the constraints are created by the leaves themselves through variations of the internal turgor pressure (see below). The major difficulty consists in modeling adequately the resulting constraints.

Going deeper into the detail, for a given mesh the strain energy $E_{\text{strain}}$ was computed according to

$$E_{\text{strain}} = \frac{E\,h}{2(1-\nu^2)} \sum_{n=1}^{M} [(1-\nu)\text{Tr}(\boldsymbol{\varepsilon}_n^2) + \nu(\text{Tr}(\boldsymbol{\varepsilon}_n))^2]\,\delta S_n\;,$$

(2.2)

where $\delta\mathcal{S}_n$ is the area of facet $n$, while the Cauchy-Green strain tensor [17] for facet $n$, $\boldsymbol{\varepsilon}_n$, writes

$$\boldsymbol{\varepsilon}_n = \frac{1}{2}(\mathbf{F}_n \cdot (\mathbf{F}_n^0)^{-1} - \mathbf{I}) \ . \tag{2.3}$$

In this equation, $\mathbf{I}$ denotes the 2×2 identity matrix, while $\mathbf{F}_n$ and $\mathbf{F}_n^0$ are the Gram matrices for facet $n$ in the strained geometry and the reference one, that is,

$$\mathbf{F}_n = \begin{pmatrix} (\mathbf{r}_{n2} - \mathbf{r}_{n1}) \cdot (\mathbf{r}_{n2} - \mathbf{r}_{n1}) & (\mathbf{r}_{n2} - \mathbf{r}_{n1}) \cdot (\mathbf{r}_{n3} - \mathbf{r}_{n1}) \\ (\mathbf{r}_{n2} - \mathbf{r}_{n1}) \cdot (\mathbf{r}_{n3} - \mathbf{r}_{n1}) & (\mathbf{r}_{n3} - \mathbf{r}_{n1}) \cdot (\mathbf{r}_{n3} - \mathbf{r}_{n1}) \end{pmatrix} \ , \tag{2.4}$$

where $\mathbf{r}_{n1}$, $\mathbf{r}_{n2}$, and $\mathbf{r}_{n3}$ describe the positions of the three vertices of the facet. On the other hand, the terms containing $\mathrm{Tr}(\mathbf{b})$ and $\mathrm{Det}(\mathbf{b})$ in Eq. (2.1) are known as the mean curvature energy and the Gaussian curvature energy, respectively. They can be rewritten in the more explicit form

$$E_{\text{curv}} = E_{\text{mean}} + E_{\text{Gauss}}$$
$$E_{\text{mean}} = \frac{E\,h^3}{24(1-\nu^2)}\int_S (c_1 + c_2 - c_1^0 - c_2^0)^2 \, dS \tag{2.5}$$
$$E_{\text{Gauss}} = -\frac{E\,h^3}{12(1+\nu)}\int_S ((c_1 - c_1^0)(c_2 - c_2^0) - \sin^2\theta(c_1^0 - c_2^0)(c_1 - c_2)) \, dS \ ,$$

where the $c_k$ and $c_k^0$ ($k$=1,2) are the local principal curvatures of the strained membrane and those of the reference geometry, respectively, and $\theta$ is the angle by which the local principal directions of the membrane have rotated with respect to those of the reference geometry. $E_{\text{mean}}$ was computed according to Eqs. (3.8)-(3.9) of Ref. [3]. In contrast, we did not use the approximation of Eq. (3.10) of Ref. [3] for estimating $E_{\text{Gauss}}$, because this approximation was derived for nearly spherical surfaces. We instead used

$$E_{\text{Gauss}} \approx -\frac{E\,h^3}{12(1+\nu)}\int_S (c_1 - c_1^0)(c_2 - c_2^0) \, dS \ , \tag{2.6}$$

which is obtained from the exact formula in Eq. (2.5) by assuming that the principal directions of curvature do not rotate during the deformation of the surface, that is $\theta = 0$. The local principal curvatures $c_1$ and $c_2$ were in turn computed from the local mean curvature $\kappa$ (estimated from Eq. (3.9) of Ref. [3]) and the Gaussian curvature $G$ (estimated from Eq. (3.12) of Ref. [3]) according to

$$c_1 = \kappa - \sqrt{\kappa^2 - G}$$
$$c_2 = \kappa + \sqrt{\kappa^2 - G} \; . \qquad (2.7)$$

For the Young's modulus, we used a value $E$=5 MPa, which is typical of parenchymatous tissues [18-20].

Finally, we considered that the motion of each vertex $j$ of a given mesh is governed by a Langevin equation with internal damping but without thermal noise, that is

$$\frac{d^2 \mathbf{r}_j}{dt^2} = -\frac{1}{m_j} \nabla E_{\text{pot}} - \gamma \sum_{k \in N_1(j)} \left( \frac{d\mathbf{r}_j}{dt} - \frac{d\mathbf{r}_k}{dt} \right) , \qquad (2.8)$$

where $k \in N_1(j)$ means that the sum runs over the vertices $k$ that are directly connected to vertex $j$, $m_j$ is the mass associated with vertex $j$ (estimated from Eq. (3.3) of Ref. [3]), and $\gamma$ is the dissipation coefficient. The damping term, which has essentially been used for *Dionaea*, is different from that in Eq. (3.13) of Ref. [3] because, for *Dionaea*, damping is essentially due to the motion of water inside the leaves and not to the friction of external water, as was assumed for *Utricularia*. Practically, Langevin equations (2.8) were integrated numerically with a leapfrog algorithm and a time step $\Delta t = 0.1$ µs.

## III - THE KINEMATIC AMPLIFICATION MECHANISM OF *Aldrovanda*

Since it is substantially simpler, the snapping mechanism of *Aldrovanda* will be considered first. Owing to the lack of detailed results in the literature, we recorded several snapping events of *Aldrovanda* traps using a high-speed camera (Fig. 1, Movie S1 [21]). The videos indicate that the 100 ms snapping motion is smooth and continuous (no sudden acceleration). The only part of the leaf that deforms noticeably lies close to the midrib, which bends inwards during closure, while the curvature of the rest of the leaf remains essentially unchanged. The videos also suggest that the midrib is pre-tensed in set conditions and that the elastic energy release, which follows trigger hairs stimulation, drives the closure of the lobes. This assumption is consistent with the findings that the motor tissues of *Aldrovanda* are located on both sides of the midrib [14] and that turgor in these cells increases during opening and decreases during snapping [22].

To confirm this scenario, we modeled the trap of *Aldrovanda* as a thin solid membrane. We found that the overall shape of the closed leaves can be satisfactorily reproduced by triangulation of the parametric surface defined by

$$
\begin{pmatrix} x \\ y \\ z \end{pmatrix} = \begin{pmatrix} \sin(\alpha)(R_0 + L\sin^2(\beta)\cos(\dfrac{\pi}{2}\dfrac{\alpha}{\alpha_{max}})) \\ \dfrac{3\sqrt{3}}{4}D\sin(\beta)\cos^2(\beta)\cos(\dfrac{\pi}{2}\dfrac{\alpha}{\alpha_{max}}) \\ R_0\cos(\alpha) - R_0 + W\sin^2(\beta)\cos(\alpha)\sqrt{\cos(\dfrac{\pi}{2}\dfrac{\alpha}{\alpha_{max}})} \end{pmatrix}, \tag{3.1}
$$

where $R_0 = 12$ mm is the radius of the circular midrib in closed configuration, $2\alpha_{max} = 20\pi/180 = 20°$ its central angle, $W = 2.6$ mm is the maximum width of the lobe (along the $z$ axis), $D = 1.54$ mm the maximum separation between the two lobes in closed configuration, and $L = 4$ mm is a shape parameter. Each point of the lobe is characterized by $\alpha \in [-\alpha_{max}, \alpha_{max}]$ and $\beta \in [0, \pi/2]$, which define its position along and perpendicular to the midrib, respectively (the midrib itself is the line obtained by setting $\beta = 0$ in Eq. (3.1)). The various mathematical functions that appear in Eq. (3.1) were adjusted so as to get a lifelike geometry. Eq. (3.1) represents only one lobe, the second lobe being obtained from the symmetry $(x, y, z) \rightarrow (x, -y, z)$. The mesh we used consists of $N$=2113 vertices and $M$=4096 triangles. The model leaf is approximately 4 mm long (along the $x$ axis) and 2.6 mm broad (along the $z$ axis). We assumed an uniform membrane thickness $h$=50 µm, which lies between the thickness of the two cell layers of the marginal zone and the three cell layers of the central zone [14].

For the closed trap described by Eq. (3.1), the midrib is an arc of a circle with curvature $c_{MR} = 1/R_0 = 83$ m$^{-1}$ (Fig. 1). Starting from this value, $c_{MR}$ was regularly decreased and the configuration with minimum elastic energy $E_{pot}$ was sought for each different curvature. We observed that the two lobes separate quite rapidly with decreasing $c_{MR}$, but $E_{pot}$ remains a smooth function thereof (Fig. 2, Movie S2 [21]). Since buckling is associated with discontinuities of $E_{pot}$ (see Sec. IV), the smoothness of the curve in Fig. 2 indicates that buckling is not involved in the snapping of *Aldrovanda* and that its opening and closure arise uniquely from cell swelling/shrinking mechanisms. They probably follow essentially the same pathway in reverse directions, although it remains also conceivable that re-opening of the lobes may in part be due to (or may be supported by) growth movements.

We furthermore tested by integrating Langevin equations without damping (that is, by setting $\gamma = 0$ in Eq. (2.8)) that the elastic relaxation time from B to A is shorter than 3 ms. This indicates that the observed 100 ms snapping time is imposed either by the rate of turgor variation along the midrib or by viscous damping due to water inside or outside the lobes.

The model therefore strongly suggests that (i) snapping in *Aldrovanda* does not involve buckling, so that *Aldrovanda* should indeed be classed in the set of swelling/shrinking dominated plants [1], and (ii) snapping is driven by the midrib and its neighbouring cellular structures, with the rest of the leaf playing no active role in the motion. It moreover points out that the unique feature that enables *Aldrovanda* to shut so fast is the *kinematic amplification* of the bending deformation of the midrib. The geometry of the leaf is indeed optimized so that a minute displacement of the midrib is sufficient to trigger a large opening of the lobes (Fig. 2).

## IV - THE LOBE BUCKLING MECHANISM OF *Dionaea*

Photographs and videos suggest that the snapping mechanism of *Dionaea* works completely different (Fig. 3, Movie S3 [21]). Indeed, the midrib does not deform during closure while, in contrast, the lobes invert their curvature from convex to concave [11]. Simulations performed with the membrane model and lifelike leaf geometries confirm that, for *Dionaea* geometry, it is not possible to perform opening/closing cycles just by deforming the midrib. This observation is also consistent with the early finding of Darwin [6], later confirmed by several authors [23,24], that the lobes consist of two distinct layers of cells and that the process of trap closure is actually driven by the difference in their behavior. More precisely, the cells at the inner surface of the lobes release a certain amount of water and shrink upon stimulation, while the cells at the external surface take up this water and expand rapidly [12,25] (as written in the Introduction), resulting in the fast snapping of the lobes. Two models, that described this mechanism in terms of bending elasticity [11,26], came to the diverging conclusions that snapping may [11] or may not [12,26] involve buckling of the lobes. However, both models are rather imprecise in the sense that they either take only average curvatures into account [11] or consider that the lobes are spherical surfaces [26], so that it is difficult to determine which conclusion is correct.

To solve this question, we also modeled the trap of *Dionaea* as a thin solid membrane. The overall shape of the closed leaves was reproduced by triangulation of the parametric surface defined by

$$\begin{pmatrix} x \\ y \\ z \end{pmatrix} = \begin{pmatrix} \sin(\alpha)(R_0 + \beta W \cos(d\alpha)) \\ 2D\beta(1-\beta)\cos(\dfrac{\pi}{2}\dfrac{\alpha}{\alpha_{max}}) \\ R_0\cos(\alpha) - R_0 + \beta W \cos(\alpha)\cos(d\alpha) \end{pmatrix}, \tag{4.1}$$

where $R_0 = 2$ cm is the radius of the circular midrib, $2\alpha_{max} = 50\pi/180 = 50°$ its central angle, $W = 1$ cm is the maximum width of the lobe (along the $z$ axis), $D = 3$ mm the maximum separation between the two lobes in closed configuration, and $d = 2$ is a shape parameter. Each point of the lobe is characterized by $\alpha \in [-\alpha_{max}, \alpha_{max}]$ and $\beta \in [0,1]$, which define its position along and perpendicular to the midrib, respectively (the midrib itself is the line obtained by setting $\beta = 0$ in Eq. (4.1)). The various mathematical functions that appear in Eq. (4.1) were adjusted so as to get a lifelike geometry. Eq. (4.1) represents only one lobe, the second lobe being obtained from the symmetry $(x, y, z) \rightarrow (x, -y, z)$. The mesh we used consists of $N$=1073 vertices and $M$=2048 triangles. The model leaf is approximately 2 cm long (along the $x$ axis) and 1 cm broad (along the $z$ axis). We assumed uniform membrane thickness $h$=400 μm.

Moreover, the fact that each lobe of *Dionaea* leaves consists of two layers of cells mechanically connected to each other was modelled by building two virtual meshes from the original one. These virtual meshes lie on both sides of the original one and their vertices are separated by $h/2$ along the normal to the surface at the original vertex, as shown in Fig. 4.

Let us assume that vector $\mathbf{r}_j = (x_j, y_j, z_j)$ defines the position of vertex $j$ of *Dionaea*'s mesh

at a certain time $t$ and that the outward normal to the surface at this vertex, $\mathbf{n}_j$, has been

computed according to Eq. (3.14) of Ref. [3]. The meshes associated with the inner and outer

layers are obtained from

$$\mathbf{r}_j^{in} = \mathbf{r}_j - \dfrac{h}{4}\mathbf{n}_j$$
$$\mathbf{r}_j^{out} = \mathbf{r}_j + \dfrac{h}{4}\mathbf{n}_j \tag{4.2}$$

(see Fig. 4). Moreover, the effect of turgor pressure is most naturally introduced in the form of a strain parameter $s$, which states that the equilibrium dimensions of the outer virtual mesh are multiplied by a factor $\sqrt{1 - s/2}$ and those of the inner one by a factor $\sqrt{1 + s/2}$ compared to the reference configuration of Eq. (4.1). More precisely, the Gram matrix in the reference geometry writes

$$\mathbf{F}_n^{0,\text{in}} = (1 + \frac{s}{2}) \times \begin{pmatrix} (\mathbf{r}_{n2}^{0,\text{in}} - \mathbf{r}_{n1}^{0,\text{in}}) \cdot (\mathbf{r}_{n2}^{0,\text{in}} - \mathbf{r}_{n1}^{0,\text{in}}) & (\mathbf{r}_{n2}^{0,\text{in}} - \mathbf{r}_{n1}^{0,\text{in}}) \cdot (\mathbf{r}_{n3}^{0,\text{in}} - \mathbf{r}_{n1}^{0,\text{in}}) \\ (\mathbf{r}_{n2}^{0,\text{in}} - \mathbf{r}_{n1}^{0,\text{in}}) \cdot (\mathbf{r}_{n3}^{0,\text{in}} - \mathbf{r}_{n1}^{0,\text{in}}) & (\mathbf{r}_{n3}^{0,\text{in}} - \mathbf{r}_{n1}^{0,\text{in}}) \cdot (\mathbf{r}_{n3}^{0,\text{in}} - \mathbf{r}_{n1}^{0,\text{in}}) \end{pmatrix}, \tag{4.3}$$

for the inner layer and

$$\mathbf{F}_n^{0,\text{out}} = (1 - \frac{s}{2}) \times \begin{pmatrix} (\mathbf{r}_{n2}^{0,\text{out}} - \mathbf{r}_{n1}^{0,\text{out}}) \cdot (\mathbf{r}_{n2}^{0,\text{out}} - \mathbf{r}_{n1}^{0,\text{out}}) & (\mathbf{r}_{n2}^{0,\text{out}} - \mathbf{r}_{n1}^{0,\text{out}}) \cdot (\mathbf{r}_{n3}^{0,\text{out}} - \mathbf{r}_{n1}^{0,\text{out}}) \\ (\mathbf{r}_{n2}^{0,\text{out}} - \mathbf{r}_{n1}^{0,\text{out}}) \cdot (\mathbf{r}_{n3}^{0,\text{out}} - \mathbf{r}_{n1}^{0,\text{out}}) & (\mathbf{r}_{n3}^{0,\text{out}} - \mathbf{r}_{n1}^{0,\text{out}}) \cdot (\mathbf{r}_{n3}^{0,\text{out}} - \mathbf{r}_{n1}^{0,\text{out}}) \end{pmatrix}, \tag{4.4}$$

for the outer layer, where the equilibrium coordinates in the right-hand sides of Eqs. (4.3) and (4.4) are obtained from Eqs. (4.1) and (4.2). Except for this modification, the elastic energy is computed for each mesh as described in Sec. II. The elastic energy of the trap is finally obtained as the average elastic energy

$$E_{\text{pot}} = \frac{1}{2} (E_{\text{pot}}(\mathbf{r}^{\text{in}}) + E_{\text{pot}}(\mathbf{r}^{\text{out}}))$$

$$\tag{4.5}$$

of membranes of thickness $h$ centred on the inner and outer meshes, respectively.

The strain parameter $s$ therefore quantifies the difference in equilibrium dimensions in the two layers caused by the pressure difference. Coupling between strain and curvature arises from the fact that, for positive (resp. negative) $s$, the membrane tends to bend inwards (resp. outwards) in order to increase (resp. decrease) the surface of the inner mesh and to decrease (resp. increase) that of the outer mesh.

Simulations performed with this model show that, in contrast to *Aldrovanda*, *Dionaea* leaves display several equilibrium configurations for each value of $s$. The variation with $s$ of the total elastic energy of some of these configurations is shown as thick lines in Fig. 5. Curves E1 and E3 correspond to concave geometries and E2 to convex ones. Turgor pressure variations are modelled by imposing a certain rate of variation of $s$ and integrating Langevin equations of the motion (Eq. (2.8)). Most of the time, the system follows the equilibrium curve on which it currently stands, but it may also jump to another curve when the current one ceases or becomes too unstable. Two such trajectories, T1 and T2, are represented as grey dotted lines in Fig. 5. Trajectory T2 models the actual closure of the trap (Movie S4 [21]). It starts at point A, which represents the open convex equilibrium configuration that is reached during the slow setting phase. Upon stimulation of a trigger hair, water flows from the inner to the outer trap lobe cell layer, leading to a decrease of $s$. The lobes tend to bend outwards, but their convex geometry opposes this trend, so that an increasing amount of *strain* energy is stored in the leaf (Fig. 6). At point B, the lobes suddenly buckle and dissipate the largest part of this energy by changing their overall curvature from convex (curve E2) to concave (curve E3). Closure is not complete at point C. It is achieved only at point D, after an additional decrease of $s$ (Fig. 5). Interestingly, the closed geometry after snapping (point D) is not identical to that before setting (point O). This is consistent with the experimental observation that *Dionaea* leaves steadily deform, owing in part to growth processes [14], during the three or four opening/closing cycles they are capable of before they wilt.

The time evolution of the average Gaussian curvature of the lobes obtained with a $0.050$ s$^{-1}$ decrease rate for the strain parameter $s$ and a dissipation coefficient $\gamma = 3 \times 10^{4}$ s$^{-1}$ is shown in Fig. 6. Corresponding experimental curves, like the one shown in Fig. 1 of Ref. [11], strongly depend on the exact shape, size and life history (number of performed snaps) of the investigated trap, so that there is no point in superposing experimental and calculated curves. Nonetheless, a one to one correspondence can easily be established between them. In particular, the buckling event leading to the ultra-fast curvature inversion of the lobes (B to C) can clearly be distinguished from the comparatively slower motions that are driven uniquely by turgor pressure variations (A to B and C to D). The value of the dissipation coefficient was precisely adjusted so as to lower the calculated closure speed down to the experimental one during the buckling phase from B to C.

Our model therefore strongly suggests that the buckling instability indeed plays a major role in the fast snapping of *Dionaea* leaves.

**V - CONCLUSION**

In this paper, we have established reliably what are the mechanisms involved in the fast snapping of two carnivorous plant sister species, namely kinematic amplification of bending deformations for *Aldrovanda vesiculosa* and buckling of the lobes for *Dionaea muscipula*. This marked difference raises interesting questions from the evolutionary point of view. Molecular systematic studies indeed suggest that the snap traps of *Aldrovanda* and *Dionaea* both derived from a common terrestrial ancestor that had sticky 'flypaper' traps [13]. One may consequently wonder why these traps are so different now and, in particular, whether this difference is due to the aqueous/terrestrial surroundings or the size of the preys [27]. Possibly, the kinematic amplification as described for *Aldrovanda* is an optimised mechanism to obtain very fast underwater snap-trap prey capture without too much water displacement, which otherwise would result in prey loss. We are currently performing experiments to answer these questions.

**Acknowledgements:** The authors would like to thank Dr. Lubomír Adamec, Třeboň, Czech Republic, for providing plant material. S.P.'s contribution to this work is supported by the German Federal Ministry of Education and Research within the funding directive BIONA: "Bionic Innovations for Sustainable Products and Technologies".

**REFERENCES**

[1] J.M. Skotheim and L. Mahadevan, Science 308, 1308 (2005)

[2] O. Vincent, C. Weisskopf, S. Poppinga, T. Masselter, T. Speck, M. Joyeux, C. Quilliet and P. Marmottant, Proc. Roy. Soc. B 278, 2909-2914 (2011)

[3] M. Joyeux, O. Vincent and P. Marmottant, Phys. Rev. E 83, 021911 (2011)

[4] O. Vincent, I Roditchev and P. Marmottant, PLoS ONE 6, e20205 (2011)

[5] J. Braam, New Phytol. 165(2), 377 (2005)

[6] C. Darwin, *Insectivorous Plants* (John Murray, London, 1875)

[7] S.E. Williams and A.B. Bennet, Science 218, 1120 (1982)

[8] B.S. Hill and G.P. Findlay, Q. Rev. Biophys. 14, 173 (1981)

[9] D. Hodick and A. Sievers, Planta 179, 32 (1989)

[10] R. Morillon, D. Liénard, M.J. Chrispeels and J.-P. Lassalles, Plant Physiol. 127, 720 (2001)

[11] Y. Forterre, J.M. Skotheim, J. Dumais and L. Mahadevan, Nature, 433, 421 (2005)

[12] A.G. Volkov, T. Adesina, V.S. Markin and E. Jovanov, Plant Physiol. 146, 694 (2008)

[13] K.M. Cameron, K.J. Wurdack and R.W. Jobson, Am. J. Bot. 89, 1503 (2002)

[14] J. Ashida, Mem. Coll. Sci., Kyoto Imp. Univ. B 9, 141 (1934)

[15] F.I. Niordson, *Shell Theory* (North-Holland, New York, 1985)

[16] S. Komura, K. Tamura and T. Kato, Eur. Phys. J. E 18, 343 (2005)

[17] A.F. Bower, *Applied mechanics of solids* (CRC Press, 2009) (ISBN 978-1439802472)

[18] H. Alizadeh and L.J. Segerlind, Appl. Eng. Agric. 13, 507 (1997)

[19] L. Mayor, R.L. Cunha and A.M. Sereno, Food Res. Int. 40, 448 (2007)

[20] K.J. Niklas, Am. J. Bot. 75, 1286 (1988)

[21] See Supplemental Material at XXX for Movies S1 (which shows a trap of *Aldrovanda vesiculosa* snapping after manual triggering), S2 (which shows a simulation of the snapping of a trap of *Aldrovanda vesiculosa*), S3 (which shows a trap of *Dionaea muscipula* snapping after manual triggering), and S4 (which shows a simulation of the snapping of a trap of *Dionaea muscipula*).

[22] T. Iijima and T. Sibaoka, Plant Cell Physiol. 24, 51 (1983)

[23] H.N. Mozingo, P. Klein, Y. Zeevi and E.R. Lewis, Am. J. Bot. 57, 593 (1970)

[24] W.R. Fagerberg and D. Allain, Am. J. Bot. 78, 647 (1991)

[25] W.H. Brown, Am. J. Bot. 3, 68 (1916)

[26] V.S. Markin, A.G. Volkov and E. Jovanov, Plant Signal. Behav. 3, 778 (2008)

[27] T.C. Gibson and D.M. Waller, New Phytol. 183, 575 (2009)

**FIGURE CAPTIONS**

**Figure 1** (color online): *Aldrovanda vesiculosa* trap in closed (A) and open (B) configurations. The pictures on the left correspond to the first and last frames of Movie S1 [21], while the simulations on the right correspond to the first and last frames of Movie S2 [21]. The midrib, which is the arc of a circle connecting points *a* and *b*, has been highlighted in red/black in the pictures showing simulation results. $c_{\mathrm{MR}}$ denotes the curvature (inverse of the radius) of this circle.

**Figure 2** (color online): Variation of $E_{\mathrm{pot}}$, $E_{\mathrm{strain}}$ and $E_{\mathrm{curv}}$ as a function of the midrib curvature $c_{\mathrm{MR}}$ for *Aldrovanda vesiculosa* (top), and variation of the lobe aperture *cd* as a function of the midrib end-to-end distance *ab* (bottom). A and B refer to the geometries shown in Fig. 1.

**Figure 3** (color online): Photographs of a trap of *Dionaea muscipula* in set conditions (left) and just after snapping (right). Note that, after closure, lobe curvature has changed.

**Figure 4** (color online): Schematic diagram showing how the inner and outer virtual meshes are obtained from the physical one for the model of *Dionaea muscipula*.

**Figure 5** (color online): Variation of $E_{\mathrm{pot}}$ as a function of *s* (left), and geometry of model *Dionaea muscipula* traps at various points of the diagram (right). Thick solid lines (E1 to E3) represent energy curves, while gray dotted lines (T1 and T2) represent trajectories. Neither E1 nor E2 were continued to lower values of *s* because the two lobes press against each other, which is easily accounted for in trajectory calculations but not in energy minimization procedures.

**Figure 6** (color online): Time evolution of the elastic energy of a *Dionaea muscipula* trap (top plot) and its mean Gaussian curvature (bottom plot) along trajectory T2 for a 0.050 s$^{-1}$ decrease rate for *s* and a dissipation coefficient $\gamma = 3 \times 10^4$ s$^{-1}$. The value of the dissipation coefficient was adjusted so as to lower the calculated closure speed down to the experimental one during the buckling phase from B to C. Points A to D refer to the geometries shown in Fig. 5.

**FIGURE 1**

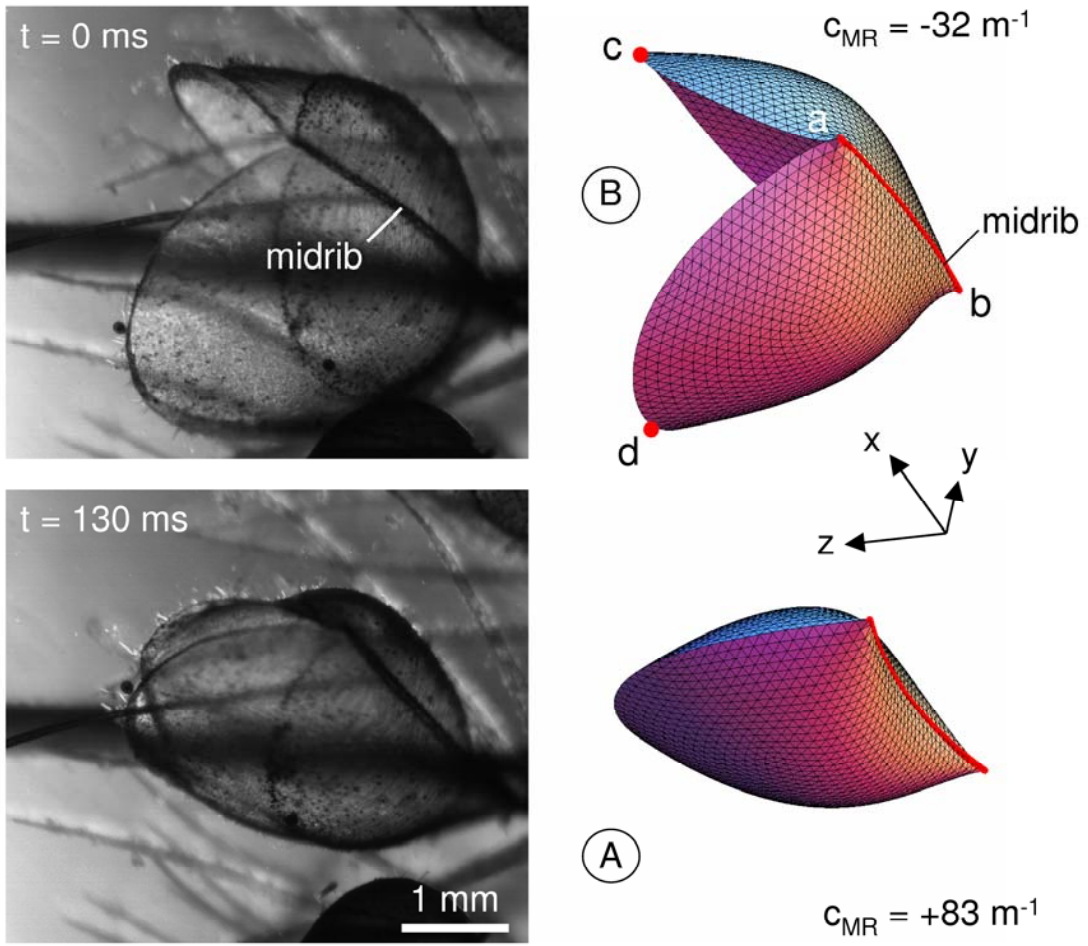

**FIGURE 2**

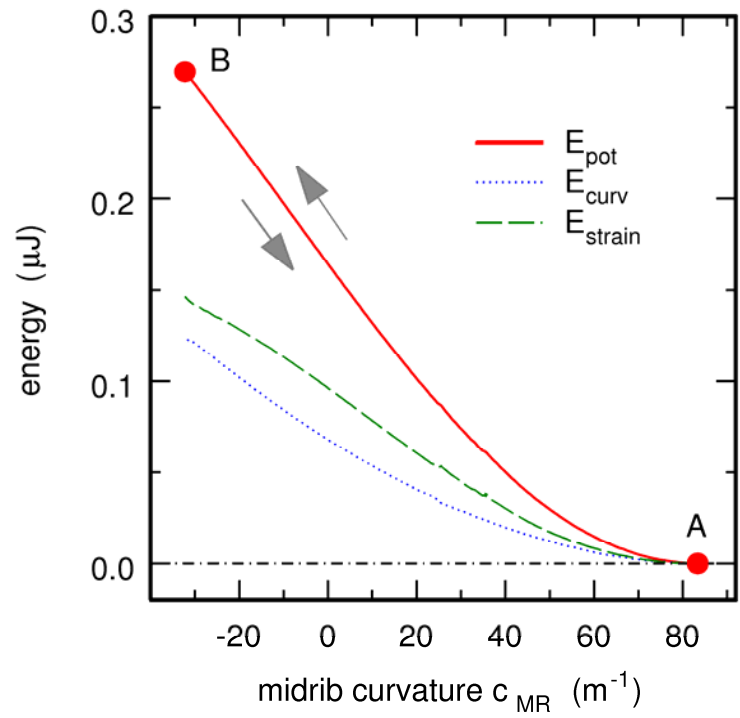

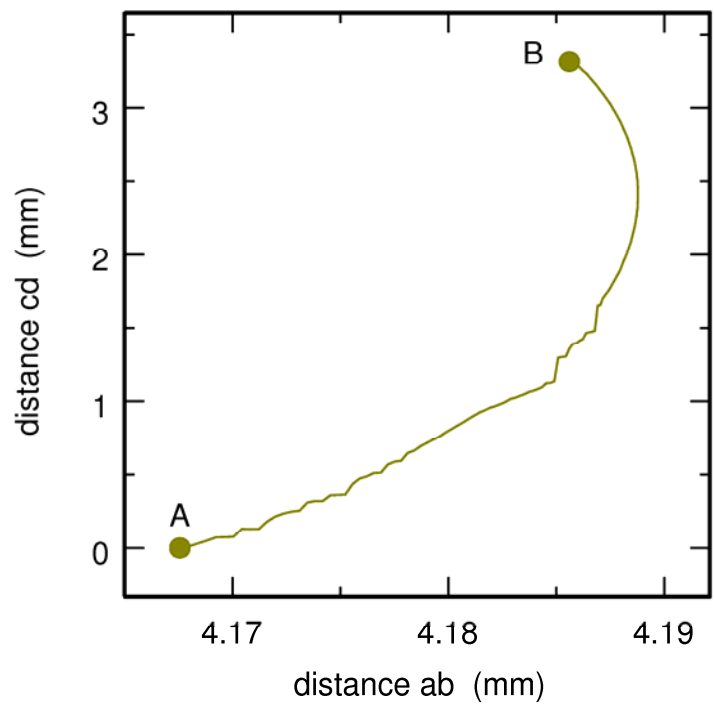

**FIGURE 3**

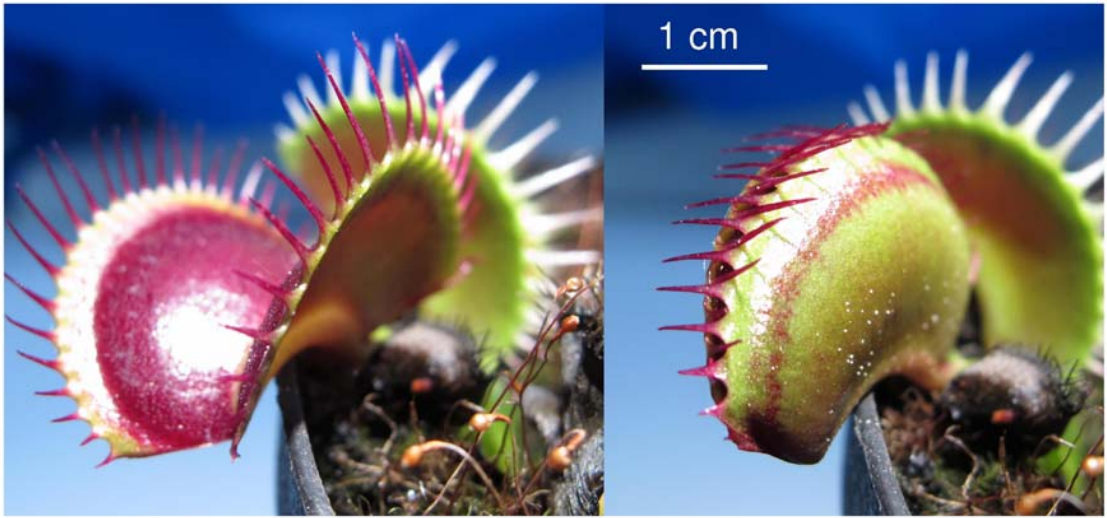

**FIGURE 4**

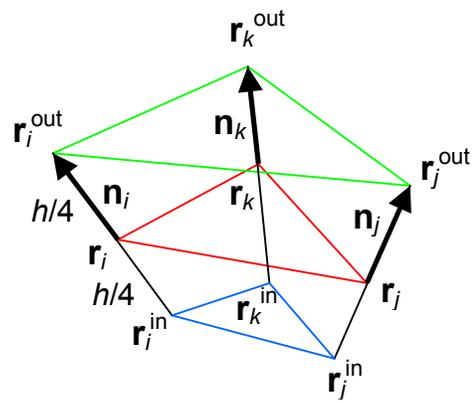

**FIGURE 5**

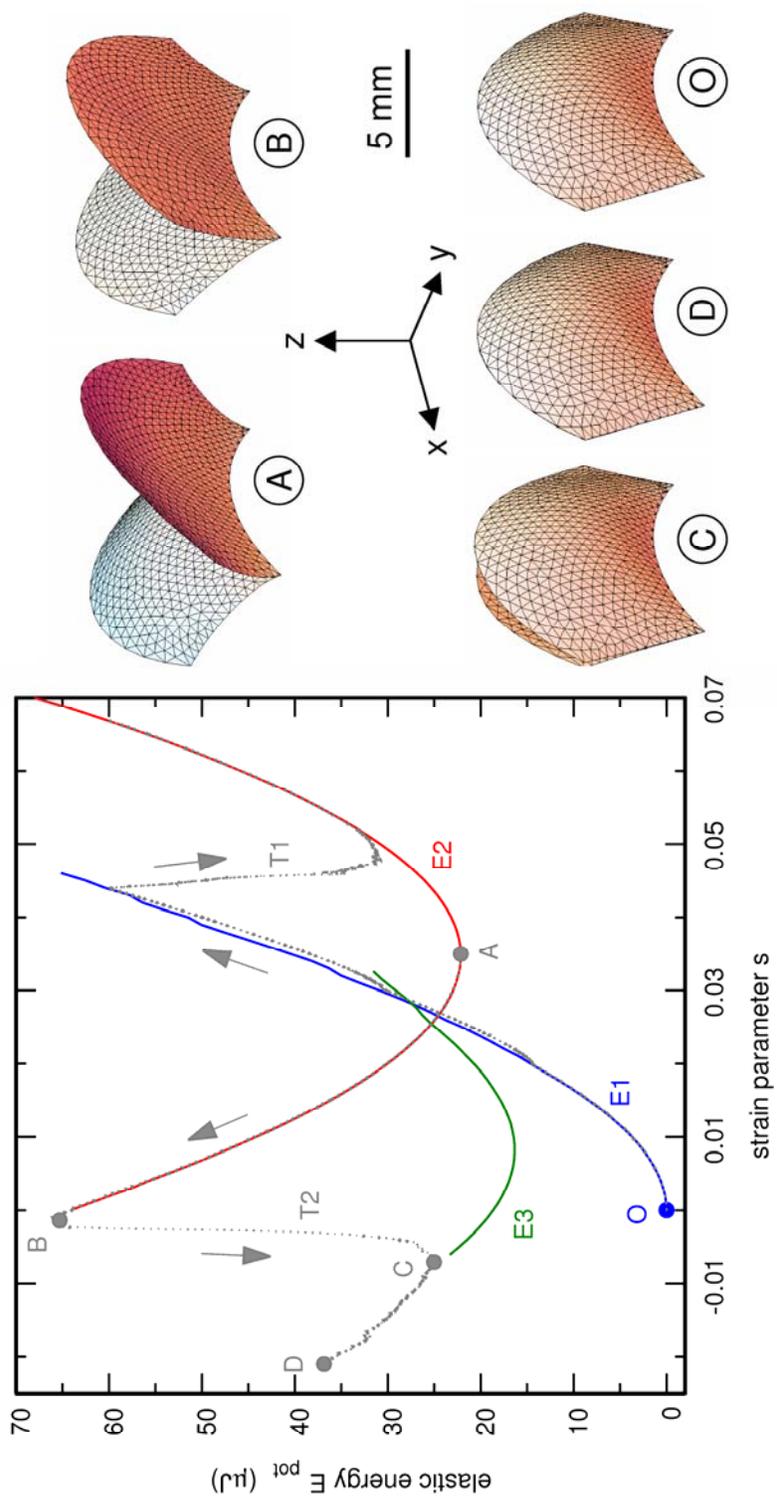

**FIGURE 6**

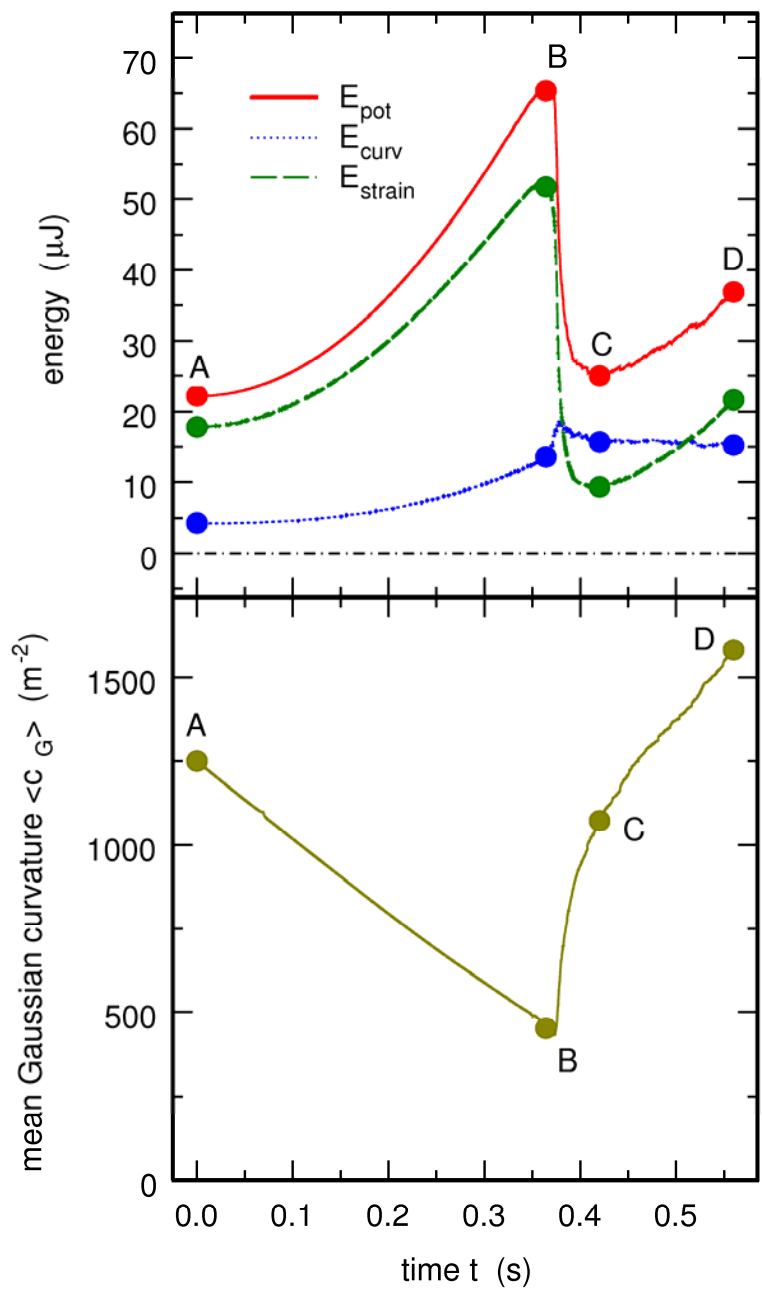